\documentclass[12pt]{article}
\usepackage{a4wide}
\usepackage{epsfig}

\newcommand{\eps}{\varepsilon}
\newcommand{\slk}{/\kern-6pt k}
\newcommand{\slp}{p\kern-5pt/}
\newcommand{\slq}{q\kern-5.5pt/}
\newcommand{\dDk}{\frac{d^Dk}{(2\pi)^D}}
\newcommand{\pfrac}[2]{\left(\frac{#1}{#2}\right)}

\begin{document}
\begin{center}
{\bf\Large Gauge dependence of the gauge boson projector}\\[24pt]
{\large E.~M.~Priidik Gallagher, Stefan Groote and Maria Naeem}\\[7pt]
Institute of Physics, University of Tartu, W.~Ostwaldi 1, 50411 Tartu, Estonia
\end{center}

\begin{abstract}
The propagator of a gauge boson, like the massless photon or the massive
vector bosons $W^\pm$ and $Z$ of the electroweak theory, can be derived in
two different ways, namely via Green's functions (semi-classical approach)
or via the vacuum expectation value of the time-ordered product of the field
operators (field theoretical approach). Comparing the semi-classical with the
field theoretical approach, the central tensorial object can be defined as
the gauge boson projector, directly related to the completeness relation for
the complete set of polarisation four-vectors. In this paper we explain the
relation for this projector to different cases of the $R_\xi$ gauge and
explain why the unitary gauge is the default gauge for massive gauge bosons.
\end{abstract}

\newpage

\section{Introduction}
As it is familiar for the scalar and Dirac propagators, the propagator of the
vector boson $V$ between two space-time locations $x$ and $y$ can be
considered as a two-point correlator, i.e.\ as the vacuum expectation value of
the time ordered product of the vector potential at these two locations,
\begin{equation}\label{vacexp}
D_V^{\mu\nu}(x-y)=\langle 0|{\cal T}\{V^\mu(x)V^\nu(y)\}|0\rangle.
\end{equation}
However, in order to get to the momentum space representation of this
propagator, one needs to use the completeness relation for the polarisation
four-vectors. This is not an easy task, as this completeness relation is not
given uniquely for a complete set of four polarisation states. As it is well
known, a massless vector boson like the photon has two polarisation states.
For a massive vector boson ($W^\pm$ or $Z$), in addition there is a
longitudinal polarisation state. However, the addition of a time-like
polarisation state is not unique and depends on the gauge we use, as we will
show in this paper. In order to get to this point, we construct the propagator
of the vector boson in a semi-classical way as Green's function obeying the
canonical equation of motion, derived as Euler--Lagrange equation from the
Lagrange density containing a gauge fixing term,
\begin{equation}
{\cal L}=-\frac12\partial_\mu V_\nu(\partial^\mu V^\nu-\partial^\nu V^\mu)
  +\frac12m_V^2 V_\mu V^\mu-\frac1{2\xi_V}(\partial_\mu V^\mu)^2,
\end{equation}
a result which will be derived in Sec.~5. $\xi_V$ is the gauge parameter in
general $R_\xi$ gauge. The solution of the Euler--Lagrange equation leads to a
propagator
\begin{equation}\label{greens}
D_V^{\mu\nu}(x-y)=\int\frac{d^4k}{(2\pi)^4}
  \frac{-iP^{\mu\nu}(k)e^{-ik(x-y)}}{k^2-m_V^2+i\epsilon},\qquad
P_V^{\mu\nu}(k):=\eta^{\mu\nu}-(1-\xi_V)\frac{k^\mu k^\nu}{k^2-\xi_Vm_V^2}
\end{equation}
with a definite second rank tensor structure $P_V^{\mu\nu}$ which we call the
gauge boson projector. $(\eta^{\mu\nu})=\mbox{diag}(1;-1,-1,-1)$ is the
Minkowski metric.

The paper is organised as follows. In Sec.~2 we introduce the gauge boson
projector. As a naive extension of the completeness relation for the
polarisation vectors fails, we offer a pragmatic solution which will be
explained in the following. In Sec.~3 we start with the Lagrange density of
the photon and explain why the solution of the corresponding Euler--Lagrange
equation needs a gauge fixing term. For a general $R_\xi$ gauge we solve the
equation for the Green's function. A recourse to historical approaches is
needed to understand the occurence of primary and secondary constraints. In
Sec.~4 the quantisation of the photon field is continued in a covariant
manner. In Sec.~5 we explain the appearance of a mass term via the Higgs
mechanism and the restriction of the gauge degrees of freedom in this case,
leading to the unitary gauge as the default setting for massive vector bosons.
In Sec.~6 we explain and give an example for the gauge independence of
physical processes. Our conclusions and outlook are found in Section~7. For
the basics we refer to
Refs.~\cite{Jackson,LandauLifshitz,Greiner,PeskinSchroeder,BohmDennerJoos}.

\section{The gauge boson projector}
The gauge boson projector as central tensorial object $P_V^{\mu\nu}(k)$ in
Eq.~(\ref{greens}) takes the simplest form $P_V^{\mu\nu}(k)=\eta^{\mu\nu}$ for
the Feynman gauge ($\xi_V=1$). For Landau gauge $\xi_V=0$ one obtains a purely
transverse projector $P_V^{\mu\nu}(k)=\eta^{\mu\nu}-k^\mu k^\nu/k^2$, and for
the unitary gauge $\xi_V\to\infty$ one has
$P_V^{\mu\nu}(k)=\eta^{\mu\nu}-k^\mu k^\nu/m_V^2$ which is transverse only on
the mass shell $k^2=m_V^2$. But why do we talk about a projector at all? A
comparison with the construction of the fermion propagator can help to explain
the conceptual approach employed in this paper.

\subsection{Construction of the fermion propagator}
As for the gauge boson propagator, there are in principle two ways to
construct the fermion propagator. As a Green's function the fermion propagator
has to solve the equation
\begin{equation}
(i\gamma^\mu\partial_\mu-m)S(x-y)=i\delta^{(4)}(x-y)
\end{equation}
equivalent to the Dirac equation $(i\gamma^\mu\partial_\mu-m)\psi(x)=0$ as the
corresponding Euler--Lagrange equation. In momentum space this equation reads
$(\slp-m)\tilde S(p)=i$ (with $\slp:=\gamma^\mu p_\mu$) which can be solved by
$\tilde S(p)=i/(\slp-m)$. Note that the inverse of the matrix $(\slp-m)$ is
well defined, since $(\slp-m)(\slp+m)=p^2-m^2$. Back to configuration space
one has
\begin{equation}
S(x-y)=\int\frac{d^4p}{(2\pi)^4}\frac{ie^{-ip(x-y)}}{p^2-m^2+i\epsilon}(\slp+m),
\end{equation}
where we have added an infinite imaginary shift $+i\epsilon$ to obtain a
Feynman propagator. On the other hand, the fermion propagator is defined again
as two-point correlator, i.e.\ as the vacuum expectation value of the
time-ordered product of the spinor and the adjoint spinor,
\begin{eqnarray}
S_{ab}(x-y)&=&\langle 0|{\cal T}\{\psi_a(x)\bar\psi_b(y)\}|0\rangle\nonumber\\
  &=&\sum_{i=1}^2\int\frac{d^3p}{(2\pi)^3}\frac1{2E(\vec p\,)}\left[
  u_i(\vec p\,)\bar u_i(\vec p\,)e^{-ip(x-y)}
  +v_i(\vec p\,)\bar v_i(\vec p\,)e^{ip(x-y)}\right]_{ab}\nonumber\\
  &=&\int\frac{d^3p}{(2\pi)^3}\frac1{2E(\vec p\,)}\left[
  (\gamma^\mu p_\mu+m)e^{-ip(x-y)}+(\gamma^\mu p_\mu-m)e^{ip(x-y)}\right]_{ab}
  \nonumber\\
  &=&(i\gamma^\mu\partial_\mu+m)_{ab}\int\frac{d^3p}{(2\pi)^3}
  \frac1{2E(\vec p\,)}\left[e^{-ip(x-y)}-e^{ip(x-y)}\right]\nonumber\\
  &=&(i\gamma^\mu\partial_\mu+m)_{ab}\int\frac{d^3p}{(2\pi)^3}
  \int\frac{dp^0}{2\pi i}\frac{-e^{-ip(x-y)}}{p^2-m^2+i\epsilon}\nonumber\\
  &=&(i\gamma^\mu\partial_\mu+m)_{ab}\int\frac{d^4p}{(2\pi)^4}
  \frac{ie^{-ip(x-y)}}{p^2-m^2+i\epsilon}\nonumber\\
  &=&\int\frac{d^4p}{(2\pi)^4}(\slp+m)_{ab}
  \frac{ie^{-ip(x-y)}}{p^2-m^2+i\epsilon},
\end{eqnarray}
where we have started with the field operators
\begin{equation}
\psi(x)=\sum_{i=1}^2\int\frac{d^3p}{(2\pi)^3}\frac1{\sqrt{2E(\vec p\,)}}
  \left(b_i(\vec p\,)u_i(\vec p\,)e^{-ipx}+\tilde b_i^\dagger(\vec p\,)
  v_i(\vec p\,)e^{ipx}\right)
\end{equation}
and $\bar\psi(x)=\psi^\dagger(x)\gamma^0$ with the only non-vanishing
antimutators
\begin{equation}
\{b_i(\vec p\,),b_j^\dagger(\vec p')\}=(2\pi)^3\delta_{ij}\delta^{(3)}
  (\vec p-\vec p'),\qquad\{\tilde b_i(\vec p\,),\tilde b_j^\dagger(\vec p')\}
  =(2\pi)^3\delta_{ij}\delta^{(3)}(\vec p-\vec p'),
\end{equation}
where we have used the completeness relations
\begin{equation}
\sum_{i=1}^2u_i(\vec p\,)\bar u_i(\vec p\,)=\gamma^\mu p_\mu+m,\qquad
\sum_{i=1}^2v_i(\vec p\,)\bar v_i(\vec p\,)=\gamma^\mu p_\mu-m,
\end{equation}
and, finally, where we have used Cauchy's theorem to write the integral in
a compact four-dimensional form. The result is quite obviously the same as
the one obtained via the Green's function. Still, one might become aware of
the central link, given by the completeness relations. A similar construction
should work also for the gauge boson propagator.

\subsection{Construction of the gauge boson propagator}
As for the quantisation of the fermion field operator we summed over the spin
polarisation states $i=1,2$ (corresponding to up and down spin), it is
natural to assume that for quantisation of the gauge boson field operator we
have to sum over the polarisations $\lambda$. Still, the (silent) assumption
that the summation runs over {\em all\/} possible (four) polarisation states
will have to be looked over again, as it will turn out. Up to that point, we
use the summation sign indexed by $\lambda$ without specifying the set of
polarisations it runs over. Therefore, starting with
\begin{equation}\label{Vquant}
V^\mu(x)=\sum_\lambda\int\frac{d^3k}{(2\pi)^3}\frac1{\sqrt{2\omega(\vec k\,)}}
  \left[\eps^\mu(\vec k,\lambda)a(\vec k,\lambda)e^{-ikx}
  +\eps^{\mu*}(\vec k,\lambda)a^\dagger(\vec k,\lambda)e^{ikx}\right]
\end{equation}
with $[a(\vec k,\lambda),a^\dagger(\vec k',\lambda')]=(2\pi)^3
\delta_{\lambda\lambda'}\delta^{(3)}(\vec k-\vec k')$ and
$\omega^2(\vec k\,)=\vec k^2+m_V^2$, the calculation of the two-point
correlator leads to
\begin{eqnarray}
\lefteqn{D_V^{\mu\nu}(x-y)\ =\ \langle 0|{\cal T}\{V^\mu(x)V^\nu(y)\}
  |0\rangle}\nonumber\\
  &=&\sum_\lambda\int\frac{d^3k}{(2\pi)^3}\frac1{2\omega(\vec k\,)}\left[
  \eps^\mu(\vec k,\lambda)\eps^{\nu*}(\vec k,\lambda)e^{-ik(x-y)}
  -\eps^\nu(\vec k,\lambda)\eps^{\mu*}(\vec k,\lambda)e^{ik(x-y)}\right].
\end{eqnarray}
However, what kind of completeness relation we can use in this case? We know
that there are at least two physical polarisation directions which are
orthogonal to each other and at the same time orthogonal to the wave vector
$\vec k$,
\begin{equation}
\vec k\cdot\vec\eps(\vec k,\lambda)=0,\qquad
\vec\eps(\vec k,\lambda)\cdot\vec\eps(\vec k,\lambda')=\delta_{\lambda\lambda'}
\end{equation}
($\lambda,\lambda'=1,2$). $\vec\eps(\vec k,1)$, $\vec\eps(\vec k,2)$ and
$\vec k/|\vec k\,|$ span an orthonormal frame. Therefore, in particular the
usual three-dimensional basis $\vec e^i$ can be expressed in this frame,
\begin{equation}
\vec e^i=\sum_{\lambda=1}^2\left(\vec e^i\cdot\vec\eps(\vec k,\lambda)\right)
  \vec\eps(\vec k,\lambda)+\frac{(\vec e^i\cdot\vec k)\vec k}{\vec k^2}
  =\sum_{\lambda=1}^2\eps^i(\vec k,\lambda)\vec\eps(\vec k,\lambda)
  +\frac{k^i\vec k}{\vec k^2}.
\end{equation}
As the usual basis is orthonormal, we conclude that
\begin{equation}
\delta^{ij}=\vec e^i\cdot\vec e^j=\sum_{\lambda=1}^2\eps^i(\vec k,\lambda)
  \eps^j(\vec k,\lambda)+\frac{k^ik^j}{\vec k^2},
\end{equation}
which can be rewritten as a first (three-dimensional) completeness relation,
\begin{equation}\label{sum2}
P_{V2}^{ij}(\vec k\,)=\sum_{\lambda=1}^2\eps^i(\vec k,\lambda)
  \eps^{j*}(\vec k,\lambda)=\delta^{ij}-\frac{k^ik^j}{\vec k^2}.
\end{equation}
Finally, considering $\vec\eps(\vec k,3):=\vec k/|\vec k\,|$ as a third
orthonomal polarisation vector, one obtains
\begin{equation}
P_{V3}^{ij}(\vec k\,)=\sum_{\lambda=1}^3\eps^i(\vec k,\lambda)
  \eps^{j*}(\vec k,\lambda)=\delta^{ij},
\end{equation}
where the complex conjugate has no effect on a real-valued basis but allows
for the generalisation for instance to a chiral basis.\footnote{We will not
make the chiral basis explicit though as we reserve $\lambda=\pm$ for
something else.} A generalisation of this completeness relation to
four-vectors (with time component set to zero) is straightforward and leads to
\begin{equation}
P_{V3}^{\mu\nu}(k)=\sum_{\lambda=1}^3\eps^\mu(k,\lambda)
  \eps^{\nu*}(k,\lambda)=\eta^\mu\eta^\nu-\eta^{\mu\nu}
\end{equation}
with $\eta^\mu=\eta^{\mu0}$. As before, an attempt can be done to switch the
non-covariant part of the right hand side to the left hand side by defining
a fourth (time-like) polarisation. However, in this simple form this attempt
fails. $\eps(k,0)=(\eta^\mu)=(1;0,0,0)$ does not give the correct sign, and
the more involved trial $\eps(k,0)=(i;0,0,0)$ is of no help here as the
product with the conjugate will remove the effect of the imaginary unit.

\subsection{The issue of dispersion}
The canonical field quantisation in Eq.~(\ref{Vquant}) is based on plane waves.
This issue becomes problematic if we consider a vector field complemented by a
gauge fixing term, leading to a nontrivial dispersion of the solution in the
case of a massive vector boson~\cite{Pavel:1997pi}. As we will see in Sec.~5,
the Proca equation can no longer be considered as a vector extension of the
Klein--Gordon equation. Instead, the mass of the vector boson depends on the
gauge parameter $\xi_V$. Accordingly, the canonical quantisation based on a
particle with fixed mass cannot be applied. However, in our approach we are
able to circumvent the problem related to the canonical quantisation by using
Green's functions. Note that Green's functions are classical and, therefore,
independent of the quantisation scheme.

\subsection{A pragmatic solution}
At this point we offer a pragmatic solution. As we know the explicit form of
the gauge boson propagator from the Green's function approach employed before,
we conclude that
\begin{equation}
P_{V4}^{\mu\nu}(k)=\sum_\lambda\eps^\mu(k,\lambda)\eps^{\nu*}(k,\lambda)
  =\eta^{\mu\nu}-(1-\xi_V)\frac{k^\mu k^\nu}{k^2-\xi_Vm_V^2}=P_V^{\mu\nu}(k).
\end{equation}
Therefore, the completeness relation depends on the gauge. The pragmatic
solution tells us that for Feynman gauge $\xi_V=1$ for instance one obtains
$\sum_\lambda\eps^\mu(k,\lambda)\eps^{\nu*}(k,\lambda)=\eta^{\mu\nu}$,
independent of whether we know which polarisations are summed over and how the
explicit polarisation vectors look like. However, we can speculate about how
these two are related to each other. We can assure ourselves that a gauge
boson on the mass shell has only vector components. In this case we obtain the
Landau projector
($\xi_V=0$)~\cite{Korner:2014bca,Berge:2015jra,Czarnecki:2018vwh}
\begin{equation}\label{sum3}
-\sum_{\lambda=1}^3\eps^\mu(k,\lambda)\eps^{\nu*}(k,\lambda)
  =\eta^{\mu\nu}-\frac{k^\mu k^\nu}{k^2}=P^{\mu\nu}_{\mathbf{1}}(k)
\end{equation}
containing only the vector component of the polarisation. Eq.~(\ref{sum3}) can
be explicitly seen in the rest frame of the massive vector boson. For
$k=(m_V;\vec 0)$ one obtains
\begin{equation}
\eta^{\mu\nu}-\frac{k^\mu k^\nu}{m_V^2}
  =-\pmatrix{0&0&0&0\cr 0&1&0&0\cr 0&0&1&0\cr 0&0&0&1\cr}
  =-\sum_{\lambda=1}^3\eps^\mu(\vec k,\lambda)\eps^{\nu*}(\vec k,\lambda)
\end{equation}
with $\eps(\vec k,1)=(0;1,0,0)$, $\eps(\vec k,2)=(0;0,1,0)$ and
$\eps(\vec k,3)=(0;0,0,1)$. If the gauge boson is offshell, it is described by
the unitary projector ($\xi_V\to\infty$), containing also a scalar
component~\cite{Berge:2015jra},
\begin{equation}\label{sum4}
\sum_{\lambda,\lambda'=0}^3\eta_{\lambda,\lambda'}\eps^\mu(k,\lambda)
  \eps^{\nu*}(k,\lambda')=\eta^{\mu\nu}-\frac{k^\mu k^\nu}{m_V^2}
  =P^{\mu\nu}_{\mathbf{1}}+\frac{k^\mu k^\nu}{k^2}F_S(k^2)
  =P^{\mu\nu}_{\mathbf{1\oplus0}}
\end{equation}
with $F_S(k^2)=1-k^2/m_V^2$ as the offshellness dominating the scalar
component. The appearance of the components $\eta_{\lambda\lambda'}$ of the
metric tensor $\eta$ in polarisation space seems to suggest that the summation
over $\lambda$ can be understood as the contraction of covariant with
contravariant components in polarisation spacetime, reserving for the
polarisation vectors the role of a tetrad between ordinary spacetime and
polarisation spacetime. This will be worked out in more detail in Sec.~4 in
case of the photon (cf.\ Eq.~(\ref{tetrad})).

\section{Green's function of the photon}
In order to investigate the relation between completeness relation and
propagator in detail, we start with the Lagrange density of the photon,
\begin{equation}\label{calLA}
{\cal L}_A=\frac12(\vec E^2-\vec B^2)=-\frac14F_{\mu\nu}F^{\mu\nu},\qquad
F_{\mu\nu}=\partial_\mu A_\nu-\partial_\nu A_\mu
\end{equation}
Containing only the self energy of the photon, the Euler--Lagrange equations
can be obtained by variation of the action integral $S_A=\int{\cal L}_Ad^4x$.
One obtains
\begin{eqnarray}
\delta S_A&=&-\frac14\int\delta F_{\mu\nu}F^{\mu\nu}d^4x
  \ =\ -\frac12\int(\partial_\mu\delta A_\nu-\partial_\nu\delta A_\mu)
  (\partial^\mu A^\nu-\partial^\nu A^\mu)d^4x\nonumber\\
  &=&-\int\partial_\mu\delta A_\nu(\partial^\mu A^\nu-\partial^\nu A^\mu)d^4x
  \ =\ \int\delta A_\nu\partial_\mu(\partial^\mu A^\nu-\partial^\nu A^\mu)d^4x,
\end{eqnarray}
where for the last step we have used integration by parts. In order to vanish
for an arbitrary variation $\delta A_\nu$ of the gauge field, one has to claim
that
\begin{equation}
\partial_\mu(\partial^\mu A^\nu-\partial^\nu A^\mu)
  =\partial^2A^\nu-\partial^\mu\partial^\nu A_\mu
  =(\partial^2\eta^{\mu\nu}-\partial^\mu\partial^\nu)A_\mu=0.
\end{equation}
However, the corresponding equation (a factor $i$ for later convenience)
\begin{equation}
(\partial^2\eta_{\mu\nu}-\partial_\mu\partial_\nu)D_A^{\mu\rho}(x)
  =i\eta^\nu_\rho\delta^{(4)}(x)
\end{equation}
for the Green's function $D_A^{\mu\rho}(x)$ cannot be solved, as the operator
$(\partial^2\eta^{\mu\nu}-\partial^\mu\partial^\nu)$ is not invertible. As
found by Faddeev and Popov in 1967, this problem turns out to be deeply
related to the gauge degree of freedom~\cite{Faddeev:1967fc}. The solution for
this problem is given by amending the Lagrange density by a gauge fixing term,
\begin{equation}\label{calLAp}
{\cal L}_{A+}=-\frac14F_{\mu\nu}F^{\mu\nu}
  -\frac1{2\xi_A}(\partial_\mu A^\mu)^2,
\end{equation}
the introduction of which can be understood on elementary level also as the
addition of a Lagrange multiplier times the square of $\partial_\mu A^\mu$,
restricting the solutions to those which satisfy the Lorenz gauge condition
$\partial_\mu A^\mu=0$ proposed exactly a century earlier~\cite{Lorenz:1867xq}.
This condition does not fix completely the gauge but eliminates the redundant
spin-$0$ component in the representation $(1/2,1/2)$ of the Lorentz group,
leaving a gauge degree of freedom $A^\mu\to A^\mu+\partial^\mu f$ with
$\partial^2f=0$. However, as the gauge field is not constrained {\it a
priori\/} but via a Lagrange multiplier, instead of a single gauge condition
one obtains a whole class of gauge conditions subsumed under the name of
$R_\xi$ gauges. For $\xi_A\to 0$ one obtains the Landau gauge classically
equivalent to the Lorenz gauge, for $\xi_A=1$ one obtains the Feynman gauge,
and for $\xi_A\to\infty$ one ends up with the unitary gauge, to name a few.

\subsection{Solution for the Green's function of the photon}
Varying the amended action functional with respect to the gauge field, in
this case one obtains
$(\partial^2\eta^{\mu\nu}-(1-\xi_A^{-1})\partial^\mu\partial^\nu)A_\mu=0$ and,
therefore,
\begin{equation}
\left(\partial^2\eta_{\mu\nu}-\left(1-\frac1{\xi_A}\right)\partial_\mu
  \partial_\nu\right)D_A^{\mu\rho}(x)=i\eta_\nu^\rho\delta^{(4)}(x)
\end{equation}
for the Green's function. This equation can be solved. In momentum space the
equation reads
\begin{equation}
-\left(k^2\eta_{\mu\nu}-\left(1-\frac1{\xi_A}\right)k_\mu k_\nu\right)
  \tilde D_A^{\mu\rho}(k)=i\eta_\nu^\rho,
\end{equation}
and by using the ansatz
$\tilde D_A^{\mu\nu}(k)=\tilde D^g\eta^{\mu\nu}+\tilde D^kk^\mu k^\nu$ one
obtains $(\xi_A-1)\tilde D^g-k^2\tilde D^k=0$ and $-k^2\tilde D^g=i$, i.e.\
\begin{equation}\label{propA}
D_A^{\mu\nu}(x)=\int\frac{d^4k}{(2\pi)^4}e^{-ikx}(\tilde D^g\eta^{\mu\nu}
  +\tilde D^kk^\mu k^\nu)=\int\frac{d^4k}{(2\pi)^4}
  \frac{-ie^{-ikx}}{k^2}\left(\eta^{\mu\nu}
  -(1-\xi_A)\frac{k^\mu k^\nu}{k^2}\right).
\end{equation}
Depending on how the convention for the poles at $k^2=0$ (i.e.\ at
$k^0=\pm\omega(\vec k)=\pm|\vec k\,|$) is set, one obtains a retarded,
advanced, or Feynman propagator (the latter not to be mixed up with the
Feynman gauge). In the following we restrict our attention to the Feynman
propagator, adding an infinitesimal imaginary shift $+i\epsilon$ to the
denominator.

\subsection{Going back to historical approaches}
Even though the solution of Faddeev and Popov allows to deal with the
calculation in a quite straightforward manner, in order to understand the
situation more deeply it is worth to have a look at older approaches. A very
valuable reference for this is the handbook of Kleinert~\cite{Kleinert:2016}
which will be used for the following argumentation.

Starting again with the free Lagrange density~(\ref{calLA}), for a canonical
field quantisation we have to obtain the Hamilton density by performing a
Legendre transformation. However, while the spatial components of the
canonical momentum are given by the components of the electric field, the time
component vanishes,
\begin{equation}
\pi^i(x)=\frac{\partial{\cal L}_A(x)}{\partial\dot A_i(x)}=-F^{0i}(x)
  =E^i(x),\qquad \pi^0(x)=\frac{\partial{\cal L}_A(x)}{\partial\dot A_0(x)}=0.
\end{equation}
According to Dirac's classification~\cite{Dirac:1947}, the property
$\pi^0(x)=0$ is a primary constraint on the canonical momentum. Using the
Euler--Lagrange equations, we get to the secondary constraint
$\nabla\vec E(\vec x,t)=0$ which is Coulomb's law for free fields.\footnote{In
case of an electric source the right hand side is replaced by $\rho(\vec x,t)$.}
The secondary constraint leads to an incompatibility for the canonical
same-time commutator
\begin{equation}
[\pi^i(\vec x,t),A^j(\vec x',t)]=i\delta^{ij}\delta^{(3)}(\vec x-\vec x').
\end{equation}
This problem can be solved by introducing a transverse modification of the
delta distribution~\cite{Kleinert:2016}. For the canonical quantisation,
$A^0(\vec x,t)$ and (via Coulomb's law) also $\nabla\vec A(\vec x,t)$ cannot
be considered as operators. Using Coulomb gauge $\nabla\vec A(\vec x,t)=0$,
one has $A^0(\vec x,t)=0$ as well, a relation between the Coulomb and axial
gauges as two examples for noncovariant gauges~\cite{Leibbrandt:1987qv}
established by Coulomb's law for free fields. One obtains
\begin{equation}
A^\mu(x)=\int\frac{d^3k}{(2\pi)^3}\frac1{\sqrt{2\omega(\vec k\,)}}
  \sum_{\lambda=1}^2\left(\eps^\mu(\vec k,\lambda)a(\vec k,\lambda)e^{-ikx}
  +\eps^{\mu*}(\vec k,\lambda)a^\dagger(\vec k,\lambda)e^{ikx}\right),
\end{equation}
where the polarisation sum runs over the two physical polarisation states
($\lambda=1,2$) only.

However, it is far from being convenient to impose noncovariant constraints
to a Lorentz-covariant quantity like the electromagnetic potential $A^\mu(x)$.
A much better choice would be the Lorenz gauge $\partial_\mu A^\mu=0$. By
using the gauge transformation of the first kind
$A^\mu\to A^\mu+\partial^\mu\lambda$, a scalar function
$\lambda(x)$ can be found so that after this transformation
$\partial_\mu A^\mu=0$ is satisfied. Still, the Lorenz gauge does not fix the
gauge degree of freedom completely. Indeed, a gauge transformation of the
second kind $A^\mu\to A^\mu+\partial^\mu f$ with $\partial^2f=0$, also called
restricted or on-shell gauge transformation, will change the vector potential
in a way that it still satisfies the Lorenz gauge constraint.
The covariant quantisation method is established by introducing a first type
of gauge-fixing term~\cite{Fermi:1932xva},
\begin{equation}
{\cal L}_{AF}={\cal L}_A+{\cal L}_{GF},\qquad
{\cal L}_{GF}=-G(x)\partial_\mu A^\mu(x)+\frac\xi2G^2(x),\qquad\xi\ge 0.
\end{equation}
In this case there is no canonical momentum for $G(x)$, and the
Euler--Lagrange equation will lead to the (secondary) constraint
$\xi G(x)=\partial_\mu A^\mu(x)$. The Euler--Lagrange equations for the
vector potential read $\partial^\mu F_{\mu\nu}(x)=\partial^2A_\nu(x)
-\partial^\mu\partial_\nu A_\mu(x)=-\partial_\nu G(x)$, and applying the
constraint one obtains
\begin{equation}
\partial^2A_\nu(x)-\left(1-\frac1\xi\right)\partial_\nu\partial_\mu A^\mu(x)=0.
\end{equation}
This is the same equation we obtain in case of the Faddeev--Popov approach.
Applying once more $\partial_\nu$, one obtains $\partial^2G(x)=0$, i.e.\
$G(x)$ is a massless Klein--Gordon field.

\section{The photon propagator}
We continue with the quantisation procedure for the photon field in covariant
form. The manifestly covariant expression for the quantised photon field is
given by
\begin{equation}
A^\mu(x)=\int\frac{d^3k}{(2\pi)^3}\frac1{\sqrt{2\omega(\vec k)}}
  \sum_{\lambda=0}^3\left(\eps^\mu(\vec k,\lambda)a(\vec k,\lambda)e^{-ikx}
  +\eps^{\mu*}(\vec k,\lambda)a^\dagger(\vec k,\lambda)e^{ikx}\right).
\end{equation}
For $\xi=1$ (Feynman gauge) we can choose momentum-independent polarisation
vectors $\eps^\mu(\lambda)=\eta^{\mu\lambda}$. Accordingly, these vectors obey
the orthogonality and completeness relations
\begin{equation}\label{orthcomp}
\eta^{\mu\nu}\eps_\mu^*(\lambda)\eps_\nu(\lambda')=\eta_{\lambda\lambda'},
  \qquad\sum_{\lambda,\lambda'}\eta_{\lambda\lambda'}\eps^\mu(\lambda)
\eps^{\nu*}(\lambda')=\eta^{\mu\nu}.
\end{equation}
Employing the apparatus of canonical quantisation, we are left with the
canonical same-time commutators $[A^\mu(\vec x,t),A^\nu(\vec x',t)]
  =[\dot A^\mu(\vec x,t),\dot A^\nu(\vec x',t)]=0$ and
\begin{equation}
[\dot A^\mu(\vec x,t),A^\nu(\vec x',t)]
  =i\eta^{\mu\nu}\delta^{(3)}(\vec x-\vec x')
\end{equation}
which are the same as if the components are independent massless
Klein--Gordon fields. However, the sign between the temporal components is
opposite to the spatial sector, resulting also in
$[a(\vec k,\lambda),a(\vec k',\lambda')]=
[a^\dagger(\vec k,\lambda),a^\dagger(\vec k',\lambda')]=0$ and
\begin{equation}
[a(\vec k,\lambda),a^\dagger(\vec k',\lambda')]=-\eta_{\lambda\lambda'}
(2\pi)^3\delta^{(3)}(\vec k-\vec k').
\end{equation}
As a consequence, states generated by applying $a^\dagger(\vec k,0)$ have a
negative norm,
\begin{equation}
\langle0|a(\vec k,0)a^\dagger(\vec k',0)|0\rangle=\langle0|[a(\vec k,0),
  a^\dagger(\vec k',0)]|0\rangle=-(2\pi)^3\delta^{(3)}(\vec k-\vec k').
\end{equation}
The only possibility to escape this problem is to amend the temporal creation
operator by one of the spatial ones. For instance, the states
$a^\dagger(\vec k,\pm)|0\rangle$ have both zero norm, as
$a^\dagger(\vec k,\pm)|0\rangle:=
\left(a^\dagger(\vec k,0)\pm a^\dagger(\vec k,3)\right)/\sqrt2$
commutes with its Hermitian conjugate.

Obviously, it is too strong to demand $D(\vec x,t)=0$ as an operator
condition, as this is in contradiction with the canonical commutation rules.
In order to guarantee the validity of the Lorenz condition $D(\vec x,t)=0$
at any time, one instead defines a physical state imposing Fermi--Dirac
subsidiary conditions~\cite{Fermi:1932xva,Dirac:1966,Heisenberg:1930xk}
\begin{equation}
D(\vec x,t)|\psi_{\rm phys}\rangle=0,\qquad
\dot D(\vec x,t)|\psi_{\rm phys}\rangle=0.
\end{equation}
resulting in $a(\vec k,-)|\psi_{\rm phys}\rangle=0$ and
$a^\dagger(\vec k,-)|\psi_{\rm phys}\rangle=0$, i.e.\ both creation and
annihilation operator annihilate the physical state. Using
$[a(\vec k,\pm),a^\dagger(\vec k',\mp)]=-(2\pi)^3\delta^{(3)}(\vec k-\vec k')$,
for the Hamilton operator one obtains
\begin{eqnarray}
H&=&-\int\frac{d^3k}{(2\pi)^3}\frac{k_0}2\sum_{\lambda,\lambda'=0}^3
  \eta_{\lambda\lambda'}{\cal N}\left\{a^\dagger(\vec k,\lambda)
  a(\vec k,\lambda')+a(\vec k,\lambda)a^\dagger(\vec k,\lambda')\right\}
  \nonumber\\
  &=&\int\frac{d^3k}{(2\pi)^3}k^0\left(\sum_{\lambda=1}^2
  a^\dagger(\vec k,\lambda)a(\vec k,\lambda)
  -a(\vec k,+)a^\dagger(\vec k,-)-a^\dagger(\vec k,+)a(\vec k,-)\right),\qquad
\end{eqnarray}
where ${\cal N}\{\cdots\}$ indicates normal ordering with respect to the
physical vacuum.\footnote{Note that in contrast to Ref.~\cite{Kleinert:2016}
we integrate over the wave vector instead of summing it. According to the
usual agreement for normal ordering, there is no contribution to the vacuum
energy soever.} Hence the subsidiary condition makes the last two terms vanish
for all physical states. For general $R_\xi$ gauges the orthogonality and
completeness relations~(\ref{orthcomp}) have to be replaced
by~\cite{Kleinert:2016}
\begin{equation}\label{tetrad}
P^{\mu\nu}(k)\eps_\mu^*(k,\lambda)\eps_\nu(k,\lambda')=\eta_{\lambda\lambda'},
  \qquad\sum_{\lambda,\lambda'}\eta_{\lambda\lambda'}\eps^\mu(k,\lambda)
\eps^{\nu*}(k,\lambda')=P^{\mu\nu}(k).
\end{equation}

\subsection{The Gupta--Bleuler quantisation}
Even though the application of both subsidiary conditions leads to the correct
physical result, the treatment of (infinite) normalisations of the states
dealt with in detail in Sec.~5.4.2 of Ref.~\cite{Kleinert:2016} is exhausting.
For processes with at least one particle it is sufficient to impose only the
first subsidiary condition
\begin{equation}\label{gupta}
a(\vec k,-)|\psi_{\rm``phys''}\rangle=0,
\end{equation}
leading to a pseudophysical state. This condition is the basis of the
Gupta--Bleuler approach to Quantum Electrodynamics~\cite{Bleuler:1950cy,%
Gupta:1949rh}. Note, however, that for a vacuum energy (for instance in
cavities) the nonphysical degrees of freedom are not completely eliminated.
In the Faddeev--Popov approach, this vacuum energy contribution will be
removed by the negative vacuum energy contribution of the Faddeev--Popov
ghosts.

As the operator in the Gupta--Bleuler subsidiary condition~(\ref{gupta})
contains only the positive-frequency part, the operator $G(x)$ is
necessarily a nonlocal operator. On the other hand side, the vacuum state
$|0_{\rm``phys''}\rangle$ has a unit norm which is an important advantage of
the Gupta--Bleuler formalism. However, the main virtue of the Gupta--Bleuler
quantisation scheme is that the photon propagator is much simpler than the
one obtained with the help of a four-dimensional (noncovariant) generalisation
of~(\ref{sum2}), namely~(\ref{propA}).

\subsection{The photon projector on the light cone}
Employing again the Green's function approach, we can get still to another
result. As in Eq.~(\ref{vacexp}), the free photon propagator is given by the
vacuum expectation value of time-ordered product of the field operators at
spacetime points $x$ and $y$,
\begin{equation}
D_A^{\mu\nu}(x-y)=\langle0|{\cal T}\{A^\mu(x)A^\nu(y)\}|0\rangle.
\end{equation}
Using the invariance of physical quantities under gauge transformations
\begin{equation}\label{gaugetra}
A^\mu(x)\to A^\mu(x)+\partial^\mu\lambda(x)
\end{equation}
with some arbitrary scalar function $\lambda(x)$, for the propagator one
obtains
\begin{eqnarray}
\lefteqn{D_A^{\mu\nu}(x-y)\ =\ \langle0|{\cal T}\{A^\mu(x)A^\nu(y)\}0\rangle
  \to}\nonumber\\[7pt]
  &=&\langle0|{\cal T}\{A^\mu(x)A^\nu(y)\}|0\rangle
  +\partial_x^\mu\langle0|{\cal T}\{\lambda(x)A^\nu(y)\}|0\rangle
  \strut\nonumber\\&&\strut\qquad
  +\partial_y^\nu\langle0|{\cal T}\{A^\mu(x)\lambda(y)\}|0\rangle
  +\partial_x^\mu\partial_y^\nu\langle0|\{\lambda(x)\lambda(y)\}|0\rangle
  \nonumber\\[7pt]
  &=&D_A^{\mu\nu}(x-y)+\partial_x^\mu D_A^\nu(x-y)+\partial_y^\nu D_A^\mu(y-x)
  +\partial_x^\mu\partial_y^\nu D_A(x-y),
\end{eqnarray}
where $D_A^\mu(x-y)=\langle0|{\cal T}\{\lambda(x)A^\nu(y)\}|0\rangle$ and
$D_A(x-y)=\langle0|{\cal T}\{\lambda(x)\lambda(y)\}0\rangle$ are mixed and
scalar propagators. Fourier transformed to momentum space, one obtains
\begin{eqnarray}
\tilde D_A^{\mu\nu}(k)&\to&\tilde D_A^{\mu\nu}(k)+k^\mu\tilde D_A^\nu(k)
  +\tilde D_A^\mu(k) k^\nu+k^\mu k^\nu\tilde D_A(k)\nonumber\\
  &=&\tilde D_A^{\mu\nu}(k)
  +k^\mu\Big(\tilde D_A^\nu(k)+\frac12k^\nu\tilde D_A(k)\Big)
  +\Big(\tilde D_A^\mu(k)+\frac12\tilde D_A(k)k^\mu\Big)k^\nu.
\end{eqnarray}
In a similar way as the gauge field is added in the Lagrange density,
replacing the partial derivative by a covariant derivative in order to be
able to absorb contributions from local phase transformations of the field
operators in transforming according to Eq.~(\ref{gaugetra}), the propagator
has to be extended in order to comply with the same
transformations~(\ref{gaugetra}). The appropriate form of the propagator
to comply with this is
\begin{equation}\label{proplight}
D_A^{\mu\nu}(x-y)=\int\frac{d^4k}{(2\pi)^4}\frac{-iP^{\mu\nu}(k)
  e^{-ikx}}{k^2+i\epsilon},\qquad
P^{\mu\nu}=\eta^{\mu\nu}-\frac12\left(k^\mu l^\nu(k)+l^\mu(k)k^\nu\right),
\end{equation}
where $l^\mu(k)$ is a four-component function of the wave vector $k$, the
explicit form of which turns out again to depend on the gauge. Taking for
instance $l^\mu(k)=k^\mu/k^2$, one ends up again with the Landau gauge, and
for $l^\mu(k)=0$ one reaches Feynman gauge. A third possibility is given by
the light cone mirror of the four-vector $k_+=(k^0;\vec k)=k$,
$l(k)=k_-/|\vec k\,|^2$ with $k_-=(k^0;-\vec k)$. Note that the four-component
object $k_-$ is {\em not\/} a covariant four-vector, called antiscalar in
Ref.~\cite{Kleinert:2016}. We call it light cone mirror of $k$. For the
``light-cone form'' of the photon projector $P^{\mu\nu}$ extracted from
Eq.~(\ref{proplight}) in this case, the two nonphysical polarisation
directions are eliminated. This can be seen with a simple calculation for
$\eps(\vec k,1)=(0;1,0,0)$, $\eps(\vec k,2)=(0;0,1,0)$ and
$k_\pm=|\vec k|(1;0,0,\pm 1)$,
\begin{equation}
\eta^{\mu\nu}-\frac1{2|\vec k\,|^2}(k_+^\mu k_-^\nu+k_-^\mu k_+^\nu)
  =-\pmatrix{0&0&0&0\cr 0&1&0&0\cr 0&0&1&0\cr 0&0&0&0\cr}
  =-\sum_{\lambda=1}^2\eps^\mu(\vec k,\lambda)\eps^{\nu*}(\vec k,\lambda).
\end{equation}
Identifying $\eps(\vec k,\pm)=k_\pm/\sqrt2|\vec k|$, one gets back to the
Fermi--Dirac (or Gupta--Bleuler) nonphysical modes, concluding that the given
combination of momentum vector $k_+$ and light cone mirror $k_-$ will
eliminate the nonphysical modes from the photon projector. 

\section{The gauge boson propagator}
As for the photon field, the vanishing of the temporal component of the
canonical momentum of the massive gauge boson is a primary constraint.
However, there is no secondary constraint. Replacing
$V^\mu(x)\to V^\mu(x)+\partial^\mu\lambda(x)$ and inserting this in the
Euler--Lagrange equation of the Lagrange density without gauge fixing
${\cal L}_V=-\frac14F_{\mu\nu}F^{\mu\nu}+\frac12m_V^2V_\mu V^\mu$,
\begin{equation}\label{ELmass}
\partial_\mu F^{\mu\nu}+m_V^2V^\nu=\left((\partial^2+m_V^2)\eta_{\mu\nu}
  -\partial_\mu\partial_\nu\right)V^\mu=0,
\end{equation}
one obtains $m_V^2\partial^\mu\lambda(x)=0$ which admits only the constant
solution $\lambda(x)=\lambda_0$. At the same time, the application of
$\partial_\nu$ to Eq.~(\ref{ELmass}) leads to $m_V^2\partial_\nu V^\nu=0$,
i.e.\ the Lorenz gauge by default. These two results are closely related to
each other as well as to the nonvanishing mass of the gauge boson. The gauge
degree of freedom is reduced by one, leaving three independent components for
the polarisation vector. Actually, for the massive gauge boson itself the
gauge fixing term is not necessary at all, as the operator in the second
expression in Eq.~(\ref{ELmass}) is invertible. This is the reason why the
gauge boson projector for the on-shell gauge boson field is given by default
by the projector in unitary gauge.

As it is convenient to consider the photon as the massless limit of a vector
boson, one has to add a gauge fixing term to the Lagrange density to allow for
a proper limit. Therefore, with the following consideration we are back to
the Faddeev--Popov method with a gauge fixing term allowing for a general
$R_\xi$ gauge.

\subsection{Goldstone bosons and mass terms}
Usually, the gauge bosons (except for the photon) obtain a mass via the
spontaneous symmetry breaking of the scalar Higgs field $\phi$ in the
framework of the electroweak Glashow--Weinberg--Salam (GWS) theory. At the
same time one obtains Goldstone bosons as ``scalar partners'' of the gauge
bosons. The masslessness of the photon is established due to the fact that
the corresponding scalar partner is the Higgs boson which ``sets the stage''
and keeps the photon from gaining a mass. A detailed outline of the Higgs
mechanism can be found e.g.\ in Refs.~\cite{PeskinSchroeder,BohmDennerJoos}.
Here we only briefly sketch the appearance of the Goldstone bosons and the
occurence of mass terms. Given the spontaneously broken Higgs field by
\begin{equation}
\psi=\frac1{\sqrt2}\pmatrix{h_1(x)+ih_2(x)\cr h_0+h_3(x)+ih_4(x)\cr},
\end{equation}
the scalar part of the Lagrange density can be expanded in the fields $h_i(x)$
to obtain
\begin{eqnarray}
{\cal L}_\phi&=&\left(D_\mu\phi(x)\right)^\dagger(D^\mu\phi(x))+\lambda h_0^2
  \phi^\dagger(x)\phi(x)-\lambda\left(\phi^\dagger(x)\phi(x)\right)^2
  \nonumber\\
  &=&\frac12\sum_{i=1}^4\left(D_\mu h_i(x)\right)\left(D^\mu h_i(x)\right)
  -\lambda h_0^2h_3(x)^2+O(h_i(x)^3).
\end{eqnarray}
The second term in this expansion gives a mass $m_H=h_0\sqrt{2\lambda}$ to the
Higgs boson field $h_3(x)$, while the masses of the gauge bosons are obtained
from the action of the covariant derivative
\begin{equation}
D_\mu=\partial_\mu-\frac{ig_1}2B_\mu-\frac{ig_2}2\vec W_\mu\vec\sigma
\end{equation}
at the constant part (proportional to $h_0$) of the Higgs field. One obtains
\begin{equation}
(D_\mu\phi)^\dagger(D^\mu\phi)=\frac{h_0^2}8\left[(g_1^2+g_2^2)Z_\mu Z^\mu
  +g_2^2(W^+_\mu W^{-\mu}+W^-_\mu W^{+\mu})\right].
\end{equation}
This has to be compared with the kinetic contributions
\begin{eqnarray}
{\cal L}_{WW}&=&-\frac14F_{\mu\nu}\left(U(1)\right)F^{\mu\nu}\left(U(1)\right)
  -\frac14\sum_{i=1}^3F^i_{\mu\nu}\left(SU(2)\right)
  F^{i\mu\nu}\left(SU(2)\right)\nonumber\\
  &=&-\frac12\partial_\mu B_\nu(\partial^\mu B^\nu-\partial^\nu B^\mu)
  -\frac12\sum_{i=1}^3\partial_\mu W^i_\nu(\partial^\mu W^{i\nu}
  -\partial^\nu W^{i\mu}).
\end{eqnarray}
With
\begin{equation}
W_\mu^\pm=\frac1{\sqrt2}(W_\mu^1\mp iW_\mu^2),\qquad
W_\mu^3=\frac{g_1A_\mu+g_2Z_\mu}{\sqrt{g_1^2+g_2^2}},\qquad
B_\mu=\frac{g_2A_\mu-g_1Z_\mu}{\sqrt{g_1^2+g_2^2}}
\end{equation}
one identifies the masses $m_A=0$, $m_Z=h_0\sqrt{g_1^2+g_2^2}/2$ and
$m_W=h_0g_2/2$.

In addition to the masses of the gauge bosons, the term
$(D_\mu\phi(x))^\dagger(D^\mu\phi(x))$ gives rise also to a mixing of
vector and scalar bosons,
\begin{eqnarray}
\lefteqn{\frac{ih_0g_2}{2\sqrt2}\partial_\mu(h_1+ih_2)W^{-\mu}
  -\frac{ih_0g_2}{2\sqrt2}\partial_\mu(h_1-ih_2)W^{+\mu}
  +\frac{h_0\sqrt{g_1^2+g_2^2}}2\partial_\mu h_4Z^\mu}\nonumber\\
  &=&im_W(\partial_\mu h_W^+)W^{-\mu}-im_W(\partial_\mu h_W^-)W^{+\mu}
  +m_Z(\partial_\mu h_Z)Z^\mu,
\end{eqnarray}
where it was logical to define $h_W^\pm:=(h_1\pm ih_2)/\sqrt2$ and $h_Z:=h_4$.
Using the property that the Lagrange density is determined only up to a total
derivative, these nonphysical mixing contributions will finally be cancelled
by appropriate additions to the gauge fixings in the gauge fixing terms
\begin{equation}
-\frac1{2\xi_A}G_A^2-\frac1{2\xi_Z}G_Z^2-\frac1{2\xi_W}G_W^\pm G_W^\mp,
\end{equation}
where
\begin{equation}\label{gaugefix}
G_A=\partial_\mu A^\mu,\qquad
G_Z=\partial_\mu Z^\mu-\xi_Zm_Zh_Z,\qquad
G_W^\pm=\partial_\mu W^{\pm\mu}\mp i\xi_Wm_Wh_W^\pm.
\end{equation}
In addition to the gauge fixing and the cancellation of the boson mixings, we
finally obtain mass terms also for the Goldstone bosons. The stage is now set
for calculating the propagators both for massive vector gauge bosons and the
corresponding scalar Goldstone bosons. As an example we deal with the $Z$
boson and the Goldstone boson field $h_Z$.

\subsection{Green's functions of massive gauge bosons}
For the $Z$ boson one obtains a contribution
\begin{equation}
{\cal L}_Z=-\frac12\partial_\mu Z_\nu(\partial^\mu Z^\nu-\partial^\nu Z^\mu)
  +\frac12m_Z^2Z_\mu Z^\mu-\frac1{2\xi_Z}(\partial_\mu Z^\mu)^2
\end{equation}
to the Lagrange density. The corresponding equation for the Green's function
reads
\begin{equation}
\left(\partial^2\eta_{\mu\nu}-\left(1-\frac1{\xi_Z}\right)\partial_\mu
  \partial_\nu+m_Z^2\eta_{\mu\nu}\right)D_Z^{\mu\rho}(x)
  =i\eta_\nu^\rho\delta^{(4)}(x),
\end{equation}
and this Proca equation is solved by
\begin{equation}
D_Z^{\mu\nu}(x)=\int\frac{d^4k}{(2\pi)^4}\frac{-ie^{-ikx}}{k^2-m_Z^2}
  \left(\eta^{\mu\nu}-(1-\xi_Z)\frac{k^\mu k^\nu}{k^2-\xi_Zm_Z^2}\right).
\end{equation}
For the Goldstone boson field $h_Z$ one obtains
\begin{equation}
{\cal L}_{h_Z}=\frac12(\partial_\mu h_Z)(\partial^\mu h_Z)
  -\frac1{2\xi_Z}\xi_Z^2m_Z^2h_Z^2,
\end{equation}
leading to the equation $-(\partial^2+\xi_Zm_Z^2)D^{h_Z}(x)=i\delta^{(4)}(x)$
for the Green's function solved by
\begin{equation}
D^{h_Z}(x)=\int\frac{d^4k}{(2\pi)^4}\frac{ie^{-ikx}}{k^2-\xi_Zm_Z^2}.
\end{equation}
Note the $\xi_Z$ dependence of the latter Green's function, also found in the
longitudinal part of the corresponding vector boson Green's function. For the
Landau gauge $\xi_Z=0$ for instance the mass dependence vanishes in these
parts. In this context it is worth noting that the classical equivalence to
the Lorenz gauge is directly seen from Eqs.~(\ref{gaugefix}). On the other
hand, while for Feynman gauge ($\xi_Z=1$) both vector and Goldstone bosons
carry a mass $m_Z$ and the propagators are quite similar, for the unitary
gauge ($\xi_Z\to\infty$) the Goldstone propagator vanishes, and for the vector
boson propagator one obtains
\begin{equation}
D_Z^{\mu\nu}(x)\Big|_{\xi_Z\to\infty}=\int\frac{d^4k}{(2\pi)^4}
  \frac{-ie^{-ikx}}{k^2-m_Z^2}\left(\eta^{\mu\nu}-\frac{k^\mu k^\nu}{m_Z^2}
  \right).
\end{equation}
This means that for unitary gauge the Higgs boson is the only scalar boson
that is propagated. This fact makes calculations using the unitary gauge
particularly attractive, as the scalar sector is mainly absent. Finally, we
obtain the same results also for the $W^\pm$ boson and collect our results
in Eq.~(\ref{greens}) in the Introduction.

\section{Gauge independence of processes}
Even though the gauge boson propagator depends on the $R_\xi$ gauge via the
gauge parameter $\xi$, this has no influence on particle processes. In order
to understand this, note that massive vector bosons (like $W^\pm$ and $Z$)
have to decay into pairs of fermions. Therefore, in exclusive processes the
vector boson line is terminated by a fermion line. To continue with the $Z$
boson, as the simplest example we can calculate a $Z$ boson propagator,
terminated ``on the left'' by a fermion line $f_1$ and ``on the right'' by a
fermion line $f_2$. For our considerations it does not matter whether for the
particular process the fermion lines constitute a fermion--antifermion pair
generated by (or annihilated to) the $Z$ boson, or whether it is a fermion (or
antifermion) which emits (or absorbs) the gauge boson. The gauge independence
of the process can be shown in each of these cases.

In momentum space the $Z$ boson propagator reads
\begin{equation}\label{Zprop}
\tilde D_Z^{\mu\nu}(k)=\frac{-i}{k^2-m_Z^2}\left(\eta^{\mu\nu}
  -(1-\xi_Z)\frac{k^\mu k^\nu}{k^2-\xi_Zm_Z^2}\right).
\end{equation}
It can be easily seen that this propagator can be decomposed into two
parts~\cite{Korner:2014bca},
\begin{equation}
\tilde D_Z^{\mu\nu}(k)=\frac{-i}{k^2-m_Z^2}\left(\eta^{\mu\nu}
  -\frac{k^\mu k^\nu}{m_Z^2}\right)-\frac{k^\mu k^\nu}{m_Z^2}
  \frac{i}{k^2-\xi_Zm_Z^2}.
\end{equation}
While the first part is the propagator in unitary gauge, the second part is
cancelled by the propagator of the neutral Goldstone boson field $h_Z$. In
order to show this, we replace the full (gauge-dependent) propagator by the
second term only, for this part of the matrix element obtaining (using the
Feynman rules from Appendix~A2 of Ref.~\cite{BohmDennerJoos})
\begin{eqnarray}\label{contr}
\lefteqn{\bar u(p'_2)ie\gamma^\mu\left(g_{f2}^-\Lambda_-+g_{f2}^+\Lambda_+
  \right)u(p_2)\left(-\frac{k_\mu k_\nu}{m_Z^2}\frac{i}{k^2-\xi_Zm_Z^2}\right)
  \bar u(p'_1)ie\gamma^\nu\left(g_{f1}^-\Lambda_-+g_{f1}^+\Lambda_+\right)
  u(p_1)}\nonumber\\
  &=&\frac{e^2}{m_Z^2}\bar u(p'_2)\slk\left(g_{f2}^-\Lambda_-
  +g_{f2}^+\Lambda_+\right)u(p_2)\frac{i}{k^2-\xi_Zm_Z^2}\bar u(p'_1)
  \slk\left(g_{f1}^-\Lambda_-+g_{f1}^+\Lambda_+\right)u(p_1)\qquad\quad
\end{eqnarray}
with $k=p_1-p'_1=p'_2-p_2$ and $\Lambda_\pm=(1\pm\gamma_5)/2$. The propagator
part is now reduced to the propagator of the neutral Goldstone boson.
Inserting the corresponding outer momentum differences for $k$ and using the
Dirac equations, one obtains
\begin{eqnarray}
\frac{e}{m_Z}\bar u(p'_2)\slk\left(g_{f2}^-\Lambda_-
  +g_{f2}^+\Lambda_+\right)u(p_2)
  &=&\frac{em_{f2}}{m_Z}(g_{f2}^--g_{f2}^+)\bar u(p'_2)\gamma_5u(p_2),
  \nonumber\\
\frac{e}{m_Z}\bar u(p'_1)\slk\left(g_{f1}^-\Lambda_-
  +g_{f1}^+\Lambda_+\right)u(p_1)
  &=&-\frac{em_{f1}}{m_Z}(g_{f1}^--g_{f1}^+)\bar u(p'_1)\gamma_5u(p_1).
\end{eqnarray}
Taking into account that
\begin{equation}
g_f^-=\frac{I_f^3-s_W^2Q_f}{s_Wc_W},\quad
g_f^+=\frac{s_WQ_f}{c_W}\quad\Rightarrow\quad
g_f^--g_f^+=\frac{I_f^3}{s_Wc_W}
\end{equation}
with $s_W=\sin\theta_W$, $c_W=\cos\theta_W$ the sine and cosine of the
Weinberg angle, $Q_f$ the electric charge (in units of the elementary charge
$e$) and $I_f^3$ the weak isospin of the fermion, the
contribution~(\ref{contr}) is indeed cancelled by the process with the $Z$
boson replaced by the neutral Goldstone boson, leaving us with the gauge
boson propagator in unitary gauge.

Note that in the 1960s and 1970s, the independence of physical processes under
gauge transformations were discussed as an equivalence theorem for point
transformations of the $S$ matrix~\cite{Chisholm:1961tha,Kamefuchi:1961sb,%
Salam:1971sp,Keck:1971ju,Kallosh:1972ap}. Also recently there are controversies
about whether physical processes including vector bosons are gauge invariant
(see e.g.\ Refs.~\cite{Wu:2017rxt,Gegelia:2018pjz}).

\begin{figure}[t]\begin{center}
\epsfig{figure=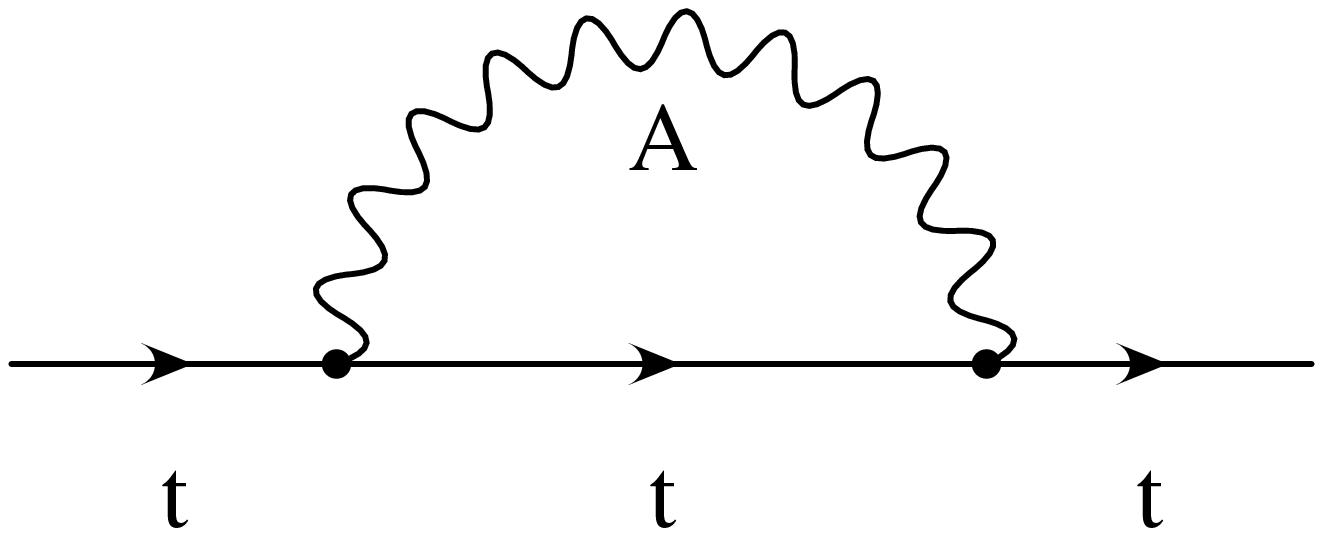,scale=0.4}\\
(a1)\\[12pt]
\epsfig{figure=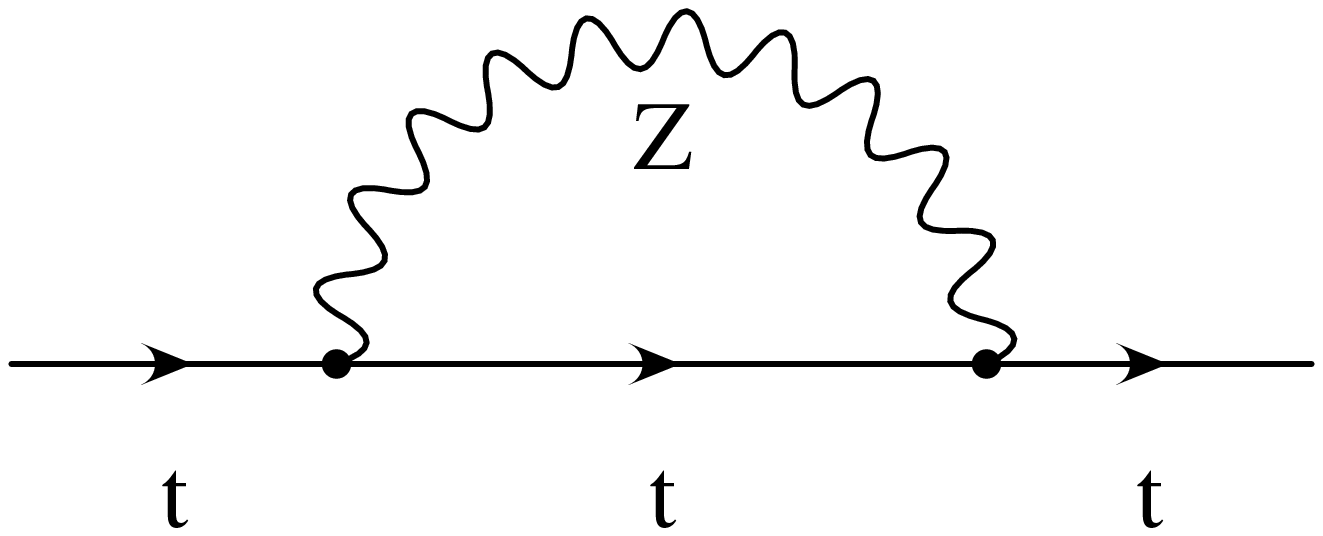,scale=0.4}\quad
\epsfig{figure=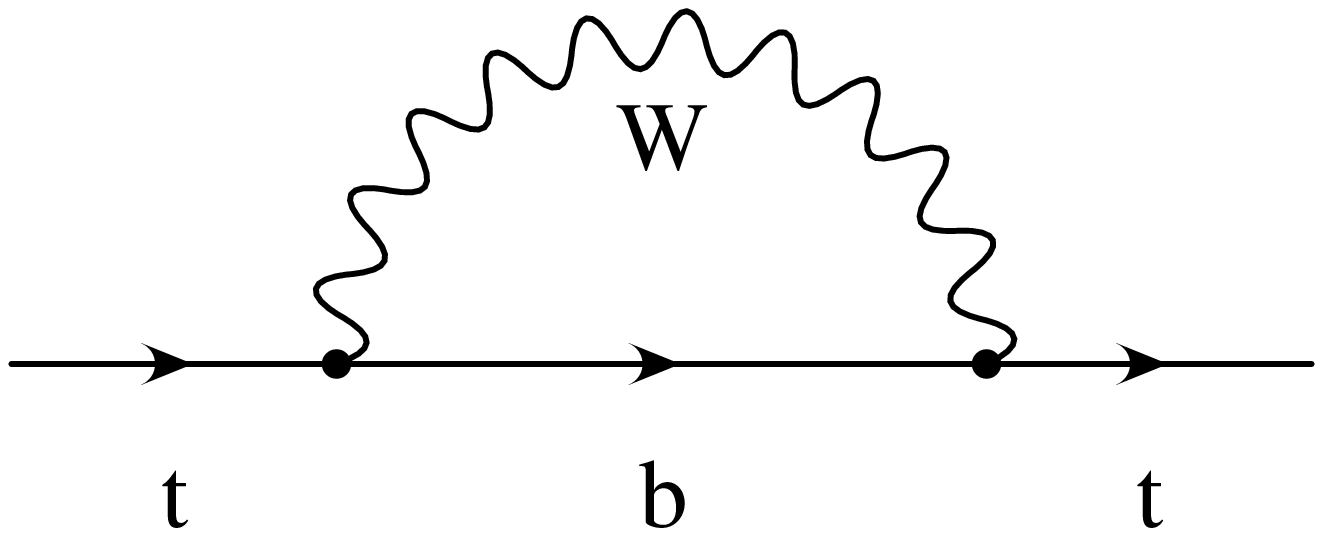,scale=0.4}\\
(a2)\kern142pt(a3)\\[12pt]
\epsfig{figure=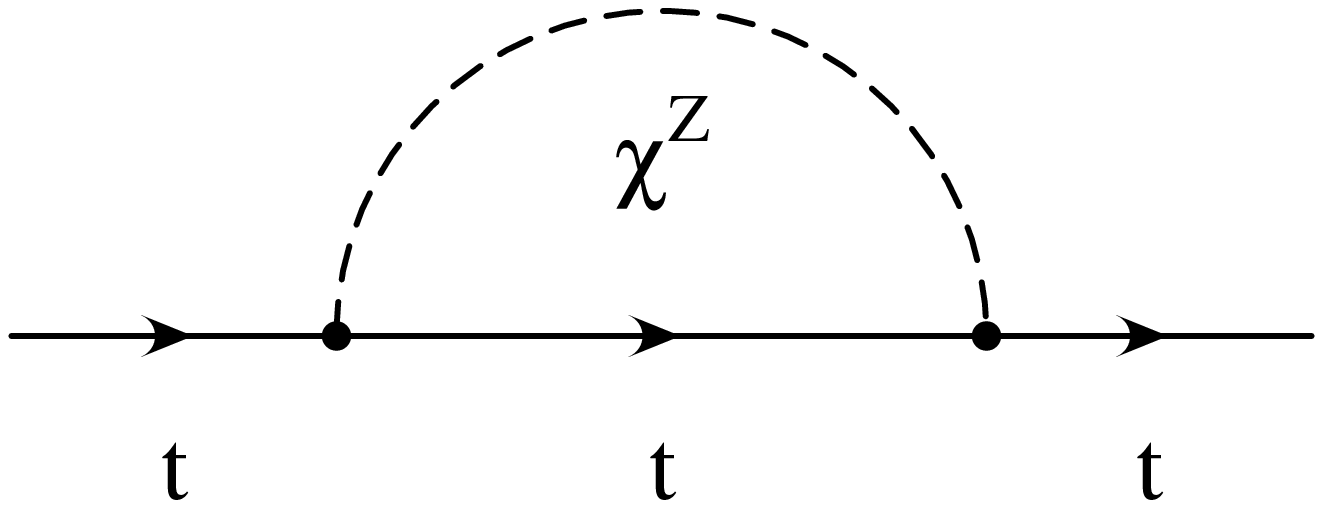,scale=0.4}\quad
\epsfig{figure=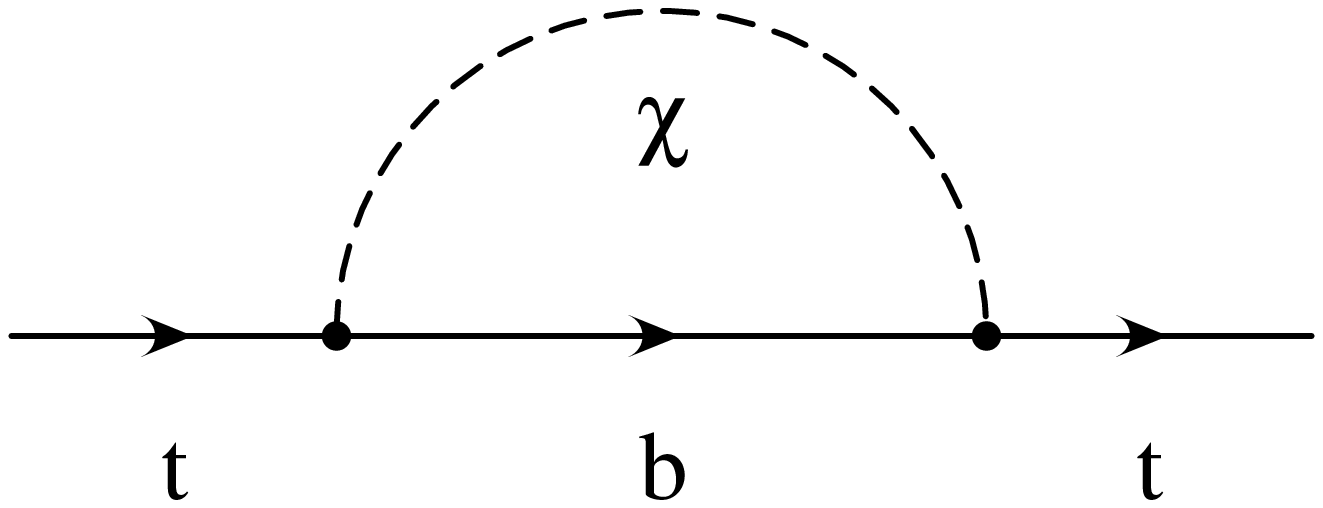,scale=0.4}\\
(b2)\kern142pt(b3)
\caption{\label{selft}top quark self energy diagrams}
\end{center}\end{figure}

\subsection{Fermion self energy contribution}
As an example for how this cancellation of the gauge dependence works out,
we calculate first order electroweak corrections to the self energy of a
fermion. As particular case we deal with the first order electroweak self
energy to the top quark. The first corrections which we denote as baseline
corrections are shown in Fig.~\ref{selft}. For the correction (a1) by a photon
one obtains
\begin{eqnarray}
i\Pi^t_{a1}&=&\int\dDk(-ieQ_t\gamma^\nu)\frac{i(\slq+\slk+m_t)}{(q+k)^2-m_t^2}
  (-ieQ_t\gamma^\mu)\frac{-i}{k^2}\left(\eta^{\mu\nu}-(1-\xi_A)\frac{k^\mu
  k^\nu}{k^2}\right)\nonumber\\
  &=&-e^2Q_t^2\int\dDk\left(\frac{\gamma^\mu(\slq+\slk+m_t)
  \gamma^\mu}{((q+k)^2-m_t^2)k^2}-(1-\xi_A)
  \frac{\slk(\slq+\slk+m_t)\slk}{((q+k)^2-m_t^2)(k^2)^2}\right).\qquad
\end{eqnarray}
Considering this correction between onshell Dirac states $\bar u(q)$ and
$u(q)$, for the second part one obtains
\begin{equation}\bar u(q)\slk(\slq+\slk+m_t)\slk u(q)
  =(2qk+k^2)\bar u(q)\slk u(q).
\end{equation}
However, using principles of dimensional regularisation, one obtains
\begin{eqnarray}
\int\dDk\frac{(2qk+k^2)\slk}{((q+k)^2-m_t^2)(k^2)^2}
  &=&\int\dDk\frac{\left((q+k)^2-m_t^2-q^2+m_t^2\right)\slk}{((q+k)^2-m_t^2)
  (k^2)^2}\nonumber\\
  &=&(-q^2+m_t^2)\int\dDk\frac\slk{((q+k)^2-m_t^2)(k^2)^2}\ =\ 0.\qquad
\end{eqnarray}
Therefore, for the correction (a1) the gauge dependence drops out, and one
obtains
\begin{equation}
i\Pi^t_{a1}=-\frac{e^2Q_t^2}{2m_t}
  \left((D-2)A(m_t)+4m_t^2B(m_t^2;m_t,m_A)\right),
\end{equation}
where $A(m)$ and $B(q^2;m_1,m_2)$ are the one- and two-point functions,
\begin{equation}
A(m)=\int\dDk\frac1{k^2-m^2},\qquad
B(q^2;m_1,m_2)=\int\dDk\frac1{\left((q+k)^2-m_1^2\right)(k^2-m_2^2)},
\end{equation}
and the photon mass $m_A$ is used as regularisator.

\begin{figure}[t]\begin{center}
\epsfig{figure=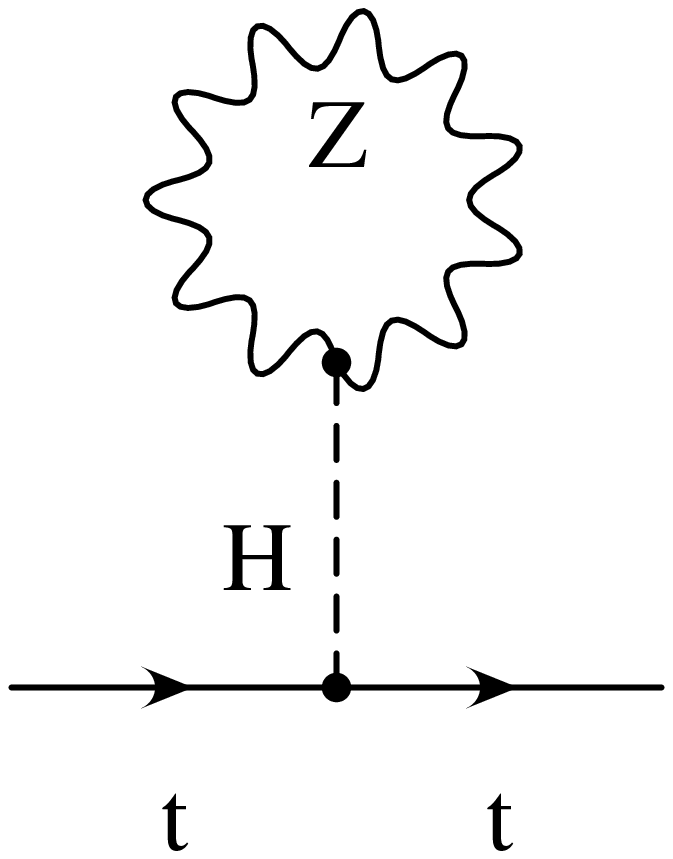,scale=0.4}\quad
\epsfig{figure=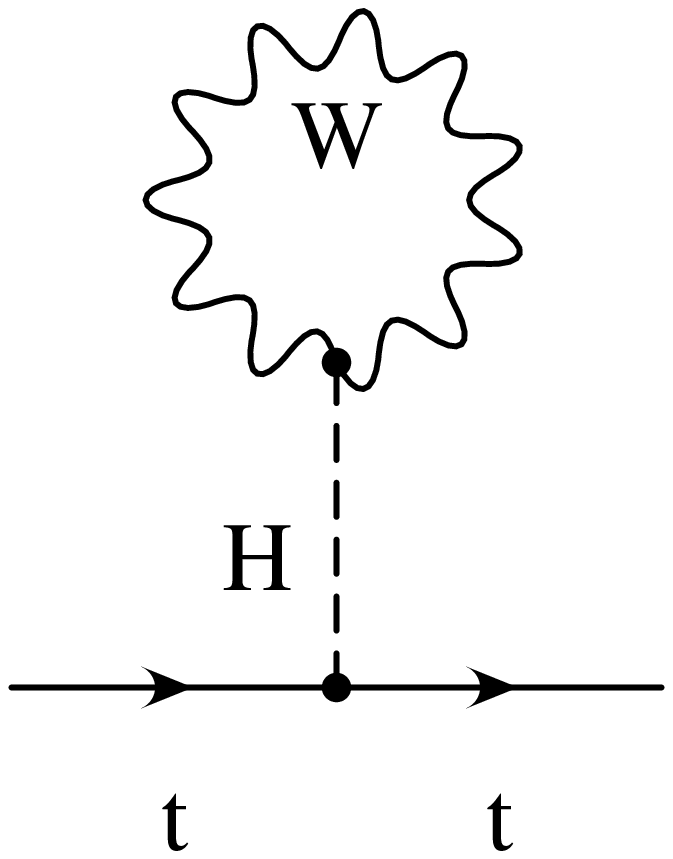,scale=0.4}\\
(c2)\kern72pt(c3)\\[12pt]
\epsfig{figure=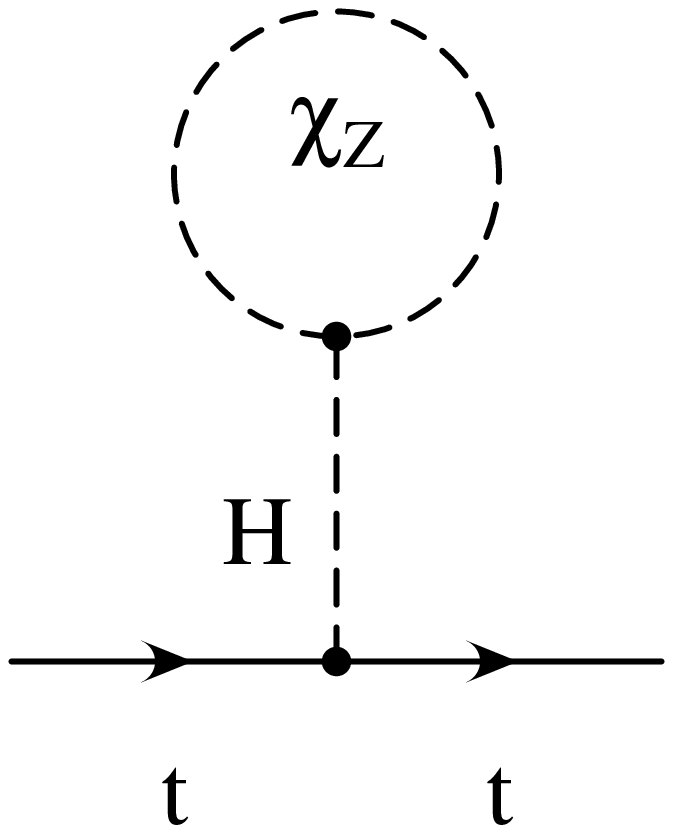,scale=0.4}\quad
\epsfig{figure=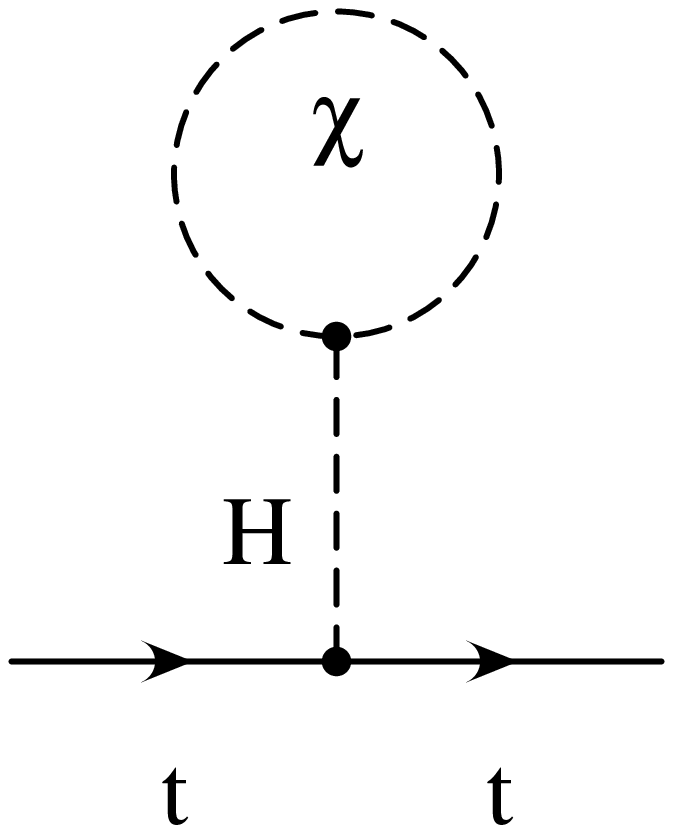,scale=0.4}\\
(d2)\kern72pt(d3)
\caption{\label{selftt}
  top quark self energy tadpole diagrams}
\end{center}\end{figure}

For the correction (a2) by the $Z$ boson the occurence of a vector boson mass
does not allow for the same conclusion. However, a first naive approach can be
tried in which the gauge dependence drops out in the sum of the corrections by
the $Z$ boson and by the corresponding Goldstone boson $\chi^Z$. In Feynman
gauge one obtains
\begin{eqnarray}
i\Pi^t_{a2}&=&\int\dDk ie\gamma^\nu\left(g_t^-\frac{1-\gamma_5}2
  +g_t^+\frac{1+\gamma_5}2\right)\frac{i(\slq+\slk+m_t)}{(q+k)^2-m_t^2}
  \times\strut\nonumber\\&&\strut
  ie\gamma^\mu\left(g_t^-\frac{1-\gamma_5}2+g_t^+\frac{1+\gamma_5}2\right)
  \frac{-ig_{\mu\nu}}{k^2-m_Z^2},\nonumber\\
i\Pi^t_{b2}&=&\int\dDk\frac{em_t}{2s_Wm_W}\gamma_5
  \frac{i(\slq+\slk+m_t)}{k^2-m_t^2}\frac{em_t}{2s_Wm_W}\gamma_5
  \frac{i}{k^2-m_Z^2}.
\end{eqnarray}
On the other hand, for unitary gauge there is no Goldstone contribution and
one stays with the correction by the $Z$ boson,
\begin{eqnarray}
i\Pi^{t\prime}_{a2}&=&\int\dDk ie\gamma^\nu
  \left(g_t^-\frac{1-\gamma_5}2+g_t^+\frac{1+\gamma_5}2\right)
  \frac{i(\slq+\slk+m_t)}{(q+k)^2-m_t^2}\times\strut\nonumber\\&&\strut
  ie\gamma^\mu\left(g_t^-\frac{1-\gamma_5}2+g_t^+\frac{1+\gamma_5}2\right)
  \frac{-i}{k^2-m_Z^2}\left(g_{\mu\nu}-\frac{k_\mu k_\nu}{m_Z^2}
  \right).\qquad
\end{eqnarray}
Looking at the difference
\begin{equation}\label{anatab2p}
i\Pi^t_{a2}+i\Pi^t_{b2}-i\Pi^{t\prime}_{a2}=\frac{e^2m_t}{8m_W^2s_W^2}A(m_Z)
\end{equation}
one realises that the difference does not vanish. However, as the difference
is proportional to the one-point function $A(m_Z)$, one might think of tadpole
contributions to be taken into account. Tadpole corrections by vector and
Goldstone bosons are shown in Fig.~\ref{selftt}. 

For Feynman gauge one obtains
\begin{eqnarray}
i\Pi^t_{c2}&=&\frac12\pfrac{-iem_t}{2s_Wm_W}\frac{i}{-m_H^2}\int\dDk
  \pfrac{iem_Wg^{\mu\nu}}{c_W^2s_W}\frac{-ig_{\mu\nu}}{k^2-m_Z^2}
  \ =\ \frac{-De^2m_t}{4c_W^2s_W^2m_H^2}\int\dDk\frac1{k^2-m_Z^2},
  \kern-28pt\nonumber\\
i\Pi^t_{d2}&=&\frac12\pfrac{-iem_t}{2s_Wm_W}\frac{i}{-m_H^2}\int\dDk
  \pfrac{-iem_H^2}{2s_Wm_W}\frac{i}{k^2-m_Z^2}
  \ =\ \frac{-e^2m_t}{8s_W^2m_W^2}\int\dDk\frac1{k^2-m_Z^2}.
  \kern-8pt\nonumber\\
\end{eqnarray}
Note the vanishing momentum square for the tadpole tail (Higgs boson). The
factor $1/2$ is a combinatorical factor due to the fact that the $Z$ boson
is its own antiparticle. As the Goldstone boson is absent for unitary gauge
(i.e.\ does not propagate), the contribution (d2) is obviously the one which
compensates the difference on the side of the Feynman gauge. However, once
again the contribution (c2) will be different for unitary gauge where one
obtains
\begin{equation}
i\Pi^{t\prime}_{c2}=\frac{-e^2m_t}{4c_W^2s_W^2m_H^2}\int\dDk\frac1{k^2-m_Z^2}
  \left(D-\frac{k^2}{m_Z^2}\right)=\frac{-(D-1)e^2m_t}{4c_W^2s_W^2m_H^2}A(m_Z).
\end{equation}

\begin{figure}[t]\begin{center}
\epsfig{figure=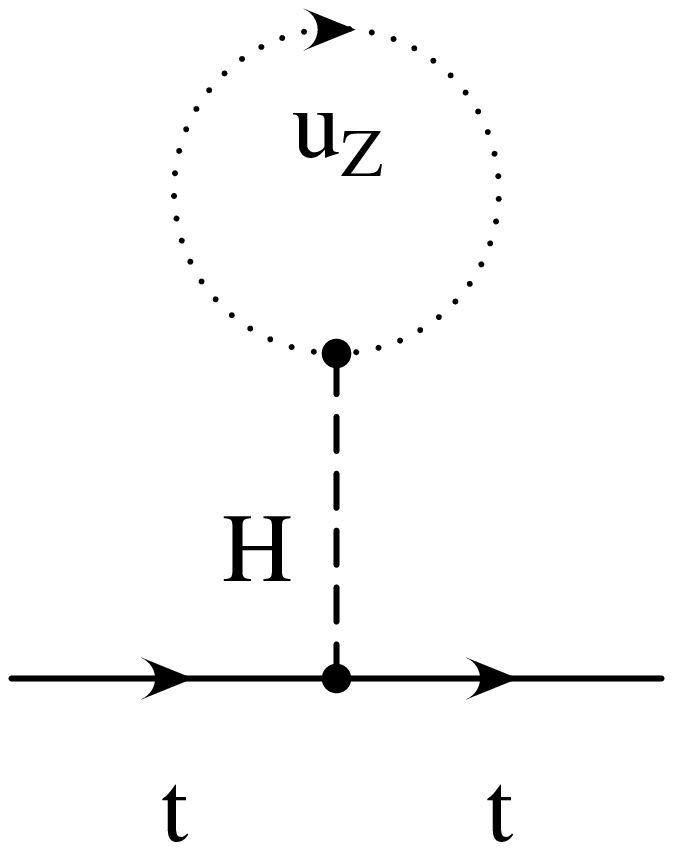,scale=0.4}\quad
\epsfig{figure=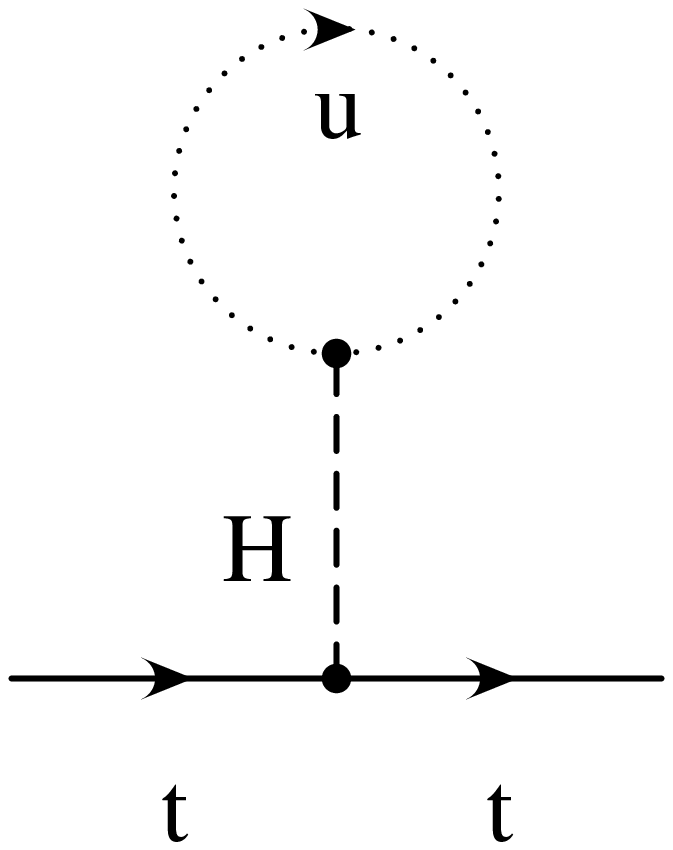,scale=0.4}\\
(e2)\kern72pt(e3)
\caption{\label{selfth}
  top quark self energy tadpole ghost diagrams}
\end{center}\end{figure}

Finally, this difference will be compensated by the corresponding ghost
contribution shown in Fig.~\ref{selfth}. For the tadpole with ghost loop
$u_Z$ one obtains
\begin{equation}
i\Pi^t_{e2}=-\pfrac{-iem_t}{2s_Wm_W}\frac{i}{-m_H^2}\int\dDk
  \pfrac{-iem_W\xi_Z}{2c_W^2s_W}\frac{i}{k^2-\xi_Zm_Z^2},
\end{equation}
where the minus sign comes from the closed ghost loop. For unitary gauge
($\xi_Z\to\infty$) the contribution remains finite. However, the dependence on
the inner momentum $k$ disappears and, therefore, there is no ghost
contribution either. On the other hand, for Feynman gauge ($\xi_Z=1$) one
obtains
\begin{equation}
i\Pi^t_{e2}=\frac{e^2m_t}{4c_W^2s_W^2m_H^2}A(m_Z).
\end{equation}
Therefore, taking into account baseline vector and Goldstone corrections
as well as tadpole vector, Goldstone and ghost corrections we obtain that the
result for Feynman gauge is the same as the one for unitary gauge.

The situation is similar in case of the corrections by $W^\pm$, $\chi^\pm$
and $u^\pm$. However, note that in this case
\begin{equation}
i\Pi^t_{a3}+i\Pi^t_{b3}-i\Pi^{t\prime}_{a3}=\frac{e^2m_t
  |V_{tb}|^2}{4m_W^2s_W^2}A(m_W).
\end{equation}
Even though we take into account only the bottom quark in the loop, the sum
in the loop has to run over all down-type quarks. Because of this fact and the
unitarity of the Cabibbo--Kobayashi--Maskawa matrix, the factor $|V_{tb}|^2$
will not appear in the final result.

\subsection{The role of unitarity}
As the parts related the two massive vector bosons $Z$ and $W^\pm$ to the self
energy of the fermion show, for the choice of unitary gauge one needs only
two instead of five contributions, namely the two contributions related to the
vector boson itself. Unitary gauge means $1/\xi=0$, i.e.\ the absence of the
gauge fixing term. Indeed, the gauge fixing term is not necessary at all if
the gauge boson carries a mass. The equation
\begin{equation}
\left(-\eta_{\mu\nu}(k^2-m_V^2)+k_\mu k_\nu\right)\tilde D_V^{\mu\rho}(k)
  =i\eta_\nu^\rho
\end{equation}
can be solved again by the ansatz
$\tilde D_V^{\mu\nu}(k)=\tilde D^g\eta^{\mu\nu}+\tilde D^kk^\mu k^\nu$, in
this case with the solution $\tilde D^g=-i/(k^2-m_V^2)$ and
$\tilde D^k=-\tilde D^g/m_V^2$, leading to the propoagator in unitary gauge,
\begin{equation}
\tilde D_V^{\mu\nu}(k)=\frac{-i}{k^2-m_V^2}\left(\eta^{\mu\nu}
  -\frac{k^\mu k^\nu}{m_V^2}\right).
\end{equation}

\section{Conclusions and Outlook}
The gauge boson projector as the central tensorial object in the propagator of
the vector gauge boson is closely related to the completeness relation for the
polarisation vectors. A generalisation of the completeness relation to
four-dimensional spacetime is proposed in a pragmatic way. Using this
approach, we could identify the polarisation vectors as tetrad fields relating
ordinary spacetime to polarisation spacetime (see Eq.~(\ref{tetrad}). While
the photon projector could be expressed by mirrors on the light cone (cf.\
Eq.~(\ref{proplight})), the projector for massive gauge bosons turned out to
be expressed in unitary gauge by default. In particular,
using the example of first order fermion self energy corrections we could show
that physical processes do not depend on the gauge degree of freedom.

From the different treatment of the massless photon and the massive vector
bosons we can draw the conclusion that the photon might not be considered as
mass zero limit of the vector boson. Indeed, at least the degree of freedoms
in this limit is not continuous. This behaviour is seen also for observables
related to the spin of particles, known as spin-flip effect (see e.g.\
Refs.~\cite{Lee:1964is,Kleiss:1986ct,Jadach:1987ws,Contopanagos:1989ga,%
Contopanagos:1992fm,Smilga:1990uq,Falk:1993tf,Korner:1993dy,Groote:1996nc,%
Groote:1997su,Dittmaier:2002nd,Groote:2009zk}). Fundamentally different Lie
group structures for massive and massless particles were investigated in
Ref.~\cite{Saar:2016jbx}, and the considerations in Ref.~\cite{Choi:2018mdd}
allow for a relation of mass and spin. Interesting enough, in combining
Refs.~\cite{Saar:2016jbx,Choi:2018mdd} a massive particle is constitued by
two massless chiral non-unitary states based on the (massless) momentum vector
and the light cone mirror of this, relating back to the light cone
representation of the photon projector. These roughly sketched relations will
be analysed in detail in a forthcoming publication.

\subsection*{Acknowledgements}
We thank J.~G.~K\"orner for useful discussions on the subject of this paper.
The research was supported by the European Regional Development Fund under
Grant No.~TK133, and by the Estonian Research Council under Grant No.~PRG356.

\end{document}